\documentclass[%
 preprint, 
 amsmath,amssymb,
 aps, physrev,
]{revtex4-2}

\usepackage{graphicx}
\usepackage{dcolumn}
\usepackage{bm}
\usepackage{natbib,hyperref}

\usepackage{siunitx}
\usepackage{gensymb}

\begin{document}

\title{\textbf{\textit{Ab initio} functional-independent calculations of the clamped Pockels tensor of tetragonal barium titanate} 
}

\author{Virginie de Mestral$^1$}
 \email{Contact author: vdemestral@ethz.ch}
\author{Lorenzo Bastonero$^2$}
\author{Michele Kotiuga$^{3,4}$}
\author{Marko Mladenovi\'{c}$^1$}
\author{Nicola Marzari$^{2,3,5}$}
\author{Mathieu Luisier$^1$}

\affiliation{%
 $^1$Integrated Systems Laboratory, ETH Zurich, Zurich, Switzerland \\
 $^2$U Bremen Excellence Chair, Bremen Center for Computational Materials Science, MAPEX Center for Materials and Processes, University of Bremen, D-28359 Bremen, Germany \\
 $^3$Theory and Simulation of Materials (THEOS), and National Centre for Computational Design and Discovery of Novel Materials (MARVEL), École Polytechnique Fédérale de Lausanne (EPFL), CH-1015 Lausanne, Switzerland \\
 $^4$Materials Design SARL, Montrouge, France \\
 $^5$Laboratory for Materials Simulations, Paul Scherrer Institut, 5232, Villigen PSI
}%

\date{\today}

\begin{abstract}
We present an \textit{ab initio} method to calculate the clamped Pockels tensor of ferroelectric materials from density-functional theory, the modern theory of polarization exploiting the electric-enthalpy functional, and automated first- and second-order finite-difference derivatives of the polarizations and the Hellmann-Feynman forces. Thanks to the functional-independent capabilities of our approach, we can determine the Pockels tensor of tetragonal barium titanate (BTO) beyond the local density approximation (LDA), with arbitrary exchange-correlation (XC) functionals, for example, PBEsol. The latter, together with RRKJ ultra-soft pseudo-potentials (PP) and a supercell exhibiting local titanium off-centering, enables us to stabilize the negative optical phonon modes encountered in tetragonal BTO when LDA and norm-conserving PP are combined. As a result, the correct value range of $r_{51}$, the largest experimental Pockels coefficient of BTO, is recovered. We also reveal that $r_{51}$ increases with decreasing titanium off-centering for this material. The lessons learned from the structural, dielectric, and vibrational investigations of BTO will be essential to design next-generation electro-optical modulators based on the Pockels effect.
\end{abstract}

\maketitle

\section{\label{sec:intro}Introduction}

Because of the growing demand for data communication, interconnects with enhanced performance in terms of speed and power consumption are actively searched for \cite{margalit2021Siphotonics_vs_electronics}. Electronic interconnects are generally limited by losses, dispersion, cross-talk, and latency effects. As such, they consume up to 50\% of the microprocessors' dynamic power. The method of choice today is to interconnect electronic devices through optical networks. These interconnects include multiple photonic integrated circuits (PICs), among which the most important are electro-optical (EO) modulators. Various PIC technologies target low-loss, low-cost, and compact integrated EO circuits, which have already found their way in data centers \cite{data_center_optical_interconnect}, high-performance computing facilities \cite{reed2023optical_modulator}, and quantum computers \cite{eltes2020cryo}. Better technologies would contribute to significant global energy savings in communication \cite{magen2004interconnect_power_dissipation}, provided that their underlying materials and components are optimally selected and designed. 
\newline 

Of particular interest is the silicon photonics solution because of its seamless integration capabilities with today's electronic circuits on a monolithic hybrid platform. However, the ultimate performance of silicon photonic platforms is hindered by technological constraints. State-of-the-art silicon-based modulators exploit the plasma-dispersion effect. They cannot exert pure phase modulation without affecting the intensity of the signal and are bound to a few tens of GHz in terms of bandwidth \cite{eltes2020cryo, hochberg2010fabless_si_photonics}. None of these issues occurs in devices that rely on the Pockels effect, which relates the change in refractive index to the modulation of an applied electric field via the Pockels tensor \cite{boyd2008nonlinear}. This linear EO effect allows for the realization of Mach-Zehnder interferometers (MZI) \cite{reider2016photonics}.

Since the Pockels effect exclusively manifests itself in non-centrosymmetric structures \cite{abel2013strongPockels}, it cannot be leveraged in silicon modulators, unless centrosymmetry is broken by design, e.g., by applying strain. Strain engineering, however, does not enable silicon to reach high-enough Pockels coefficients. Practically, lithium niobate (LNO) has established itself over recent decades as the standard material to exploit the Pockels effect in silicon PICs \cite{wooten2000LNOreview}. With a Pockels coefficient of $\sim30$ \unit{pm/V}, it is 20 times more sensitive to the electric field modulation than strained silicon \cite{chmielak2011pockels_strainedSi}. This difference is crucial as a larger Pockels coefficient entails a shorter MZI interaction length, facilitating denser silicon PIC packing and improving device miniaturization \cite{winiger2024exp_pockels}. However, LNO cannot be grown epitaxially directly on silicon due to lattice mismatches, thus requiring a multi-step layer-bonding approach that complicates large-scale manufacturing \cite{chen2012Si_multi_layer_bonding}.

A sizable Pockels coefficient combined with a monolithic integration of heterogeneous materials onto silicon PICs is highly desirable to produce next-generation ultra-fast EO modulators while keeping fabrication costs low. In contrast to LNO, BTO can be monolithically integrated onto large silicon wafers and exhibits a large Pockels coefficient $r_{51}$ peaking at 730 $\pm$ 150 \unit{pm/V} when clamped, i.e., a value that is 25 to 30 times higher than in LNO \cite{abel2019largePockels}. Indeed, integrated BTO thin-films are expected to be physically clamped within a silicon cladding. The $r_{51}$ coefficient of BTO is of greatest technological relevance, all other coefficients of the Pockels tensor being at least one order of magnitude smaller \cite{zgonik1994BTOprop}. Furthermore, BTO shows excellent EO properties and high quality crystals can be grown, both as a deposited thin-film \cite{eltes2017crystal_quality} and embedded into functionalized devices structures \cite{abel2019largePockels}. Although other high-bandwidth competitive materials exist, such as silicon-organic hybrids \cite{leuthold2013silicon_organic} and lead zirconate titanate \cite{alexander2018PZT_Pockels}, none of them is chemically and thermally as stable as BTO, or can be directly integrated into silicon PICs.

BTO is a ferrolectric perovskite. It is non-centrosymmetric in all its low-temperature phases, which is a prerequisite to observe the Pockels effect. It has a tetragonal \textit{P4mm} symmetry at room temperature, and possesses a macroscopic spontaneous polarization aligned with its $c$-axis, along the $[001]$ direction. As BTO is cooled down to 0 K, it successively becomes orthorhombic, then rhombohedral, pivoting its polarization direction to $[011]$ and $[111]$, respectively. Conversely, if it is heated above its Curie temperature of $\sim$130\unit{\degree C} \cite{acosta2017batio3_review}, it transforms into a cubic paraelectric lattice, becomes centrosymmetric, losing its Pockels properties. Being able to predict the EO characteristics of this material and, therefore, guiding the on-going experimental activity \cite{hamze2020design_rules_Pockels} is thus a goal of utmost importance.

In this work, we present a novel computational framework to calculate the clamped Pockels tensor of tetragonal BTO for any XC-functional. The unclamped Pockels tensor can be built upon the clamped one by adding a combined piezoelectric and elasto-optic contribution \cite{veithen2005nonlinear_DFPT}. We did not include this contribution in our framework because in thin-film configurations, as used experimentally \cite{abel2019largePockels}, it is negligible. Hence, there only remain electronic and optical phonon contributions. Moreover, previous works showed that the large clamped Pockels response of BTO is dominated by the interaction of lattice dynamics and electronic structures, especially by two low-energy, soft optical phonon modes and their associated large Raman susceptibilities \cite{hamze2020design_rules_Pockels, fontana1994pockels_&_dielectric_relaxation, veithen2004Pockels_soft_mode, kim2023Pockels_tetra_BTO}. 

Because of these complex interplays, \textit{ab initio} simulation methods constitute the most suitable tools to model the Pockels effect. Numerous freely available first principles codes possess lattice vibrational and non-linear optical capabilities. However, few of them offer a direct pipeline for Pockels computation, and if so, to the best of our knowledge, they operate with a restricted choice of XC-functionals. For instance, a well-established method available in the ABINIT software \cite{veithen2005nonlinear_DFPT, gonze2020abinit} relies on density-functional perturbation theory (DFPT) and the $2n+1$ theorem \cite{baroni2001DFPT} to calculate Pockels tensors. Its implementation, however, is restricted to the LDA XC-functional~\cite{veithen2005nonlinear_DFPT}. In~\cite{veithen2005nonlinear_DFPT}, the $r_{51}$ coefficient of tetragonal BTO could not be calculated due the presence of unstable soft phonon modes. To circumvent this issue, we propose an automated, functional-independent method based on standard DFT, finite displacements, and finite electric fields through the modern theory of polarization and the electric-enthalpy functional \cite{resta1994theoryPolarization}. The stabilization of the soft phonons further requires the use of a $2\times 2\times 1$ BTO supercell with local titanium off-centering along its $\langle$110$\rangle$ axes \cite{kotiuga_2022_privatecomm}, as discussed in Section \ref{chap:resI}.

In practice, we implement the clamped Pockels formalism from \cite{johnston1970LOTO_scattering_nonlinear} in the Vibroscopy software \cite{bastonero2024Vibroscopy}, which is an automated finite-difference workflow \cite{huber2020aiida, uhrin2021aiida_workflows}. It computes derivatives of Hellmann-Feynman forces and polarizations, exploiting the Quantum ESPRESSO plane-wave DFT code \cite{giannozzi2009quantum_espresso, giannozzi2017quantum_espresso, giannozzi2020quantum_espresso} and the supercell-based Phonopy code \cite{phonopy-phono3py-JPSJ}. Our functional-independent approach constitutes a comprehensive framework for predictive calculations of real Pockels-active materials, not only in their ideal form, but also in the presence of strain or defects.

The paper is organized as follows. First, we review the theoretical formalism of the clamped Pockels tensor and describe our methodology in Section \ref{chap:theory&method}. In Section \ref{chap:resI}, we show how the zone-center negative phonon modes arising from the negative curvature of the Born-Oppenheimer surface associated with soft modes \cite{ghosez1998negative_curvature_phonons} can be stabilized and turned positive. This investigation reveals that the soft modes are at the origin of the large Pockels coefficient of tetragonal BTO, and that a careful choice of XC-functional is essential to represent the physics of this material. We then present the clamped Pockels tensor obtained with our methodology in Section \ref{chap:resII} and compare it to previous works \cite{veithen2005nonlinear_DFPT, kim2023Pockels_tetra_BTO}. Lastly, we unravel the structural dependence of local titanium off-centering on the Pockels tensor in Section \ref{chap:resIV}, in particular on the soft phonon-mediated $r_{51}$ coefficient. We find that $r_{51}$ increases with decreasing titanium off-centering. Conclusions are drawn in Section \ref{chap:conclusion}.

\section{\label{chap:theory&method}Physical model}

The Pockels effect describes the change in refractive index of a non-linear crystal upon modulation of an applied electric field \cite{veithen2005nonlinear_DFPT} 
\begin{equation}
    \Delta \Bigg( \frac{1}{n_{ij}^2} \Bigg) = \Delta(\varepsilon^{-1})_{ij} = \sum_{k = 1}^3 r_{ijk} \mathcal{E}_k,
    \label{equ:method:Pockels_def}
\end{equation}
where $\mathcal{E}_k$ is the electric field, $(\varepsilon^{-1})_{ij}$ the inverse of the high-frequency dielectric tensor, $n_{ij}$ the refractive index tensor, and $r_{ijk}$ the linear EO Pockels tensor. Indices $i$, $j$, and $k$ refer to the polarization of the incoming light, outgoing light, and modulating electric field, respectively. From Eq.~(\ref{equ:method:Pockels_def}), $r_{ijk}$ can be defined as
\begin{equation}
    r_{ijk} = -\sum_{m,n=1}^3 (\varepsilon^{-1})_{im} \frac{d \varepsilon_{mn}}{d \mathcal{E}_k} (\varepsilon^{-1})_{nj}.    \label{equ:method:Pockels_E_t_strain}
\end{equation}
The dielectric tensor $\varepsilon$ is a function of the electric field $\mathcal{E}$, atomic positions $\boldsymbol{R}$, and lattice strains $\eta$. In this work, we neglect the strain-dependent contribution for the two following reasons. Firstly, the frequency of the electric field modulation typically exceeds lattice resonance, which suppresses piezoelectrically-induced acoustic strains. Secondly, the BTO thin-films of interest are physically clamped in a silicon cladding. We can thus expand the total derivative of $\varepsilon_{mn}$ in terms of the electric field and the atomic displacements $\tau$ at fixed lattice strain $\eta_0$, giving rise to an electronic (fixed atomic positions $\boldsymbol{R}_0$) and ionic response, as proposed in \cite{veithen2005nonlinear_DFPT, johnston1970LOTO_scattering_nonlinear}
\begin{equation}
    \frac{d \varepsilon_{mn}(\boldsymbol{\mathcal{E}}, \textbf{R}, \eta_0)}{d \mathcal{E}_k} = 
    \underbrace{\frac{\partial \varepsilon_{mn}(\boldsymbol{\mathcal{E}}, \textbf{R}_0, \eta_0)}{\partial \mathcal{E}_k}}_{\text{electronic}}
    + \underbrace{\sum_{\kappa, \beta} \frac{\partial \varepsilon_{mn}(\boldsymbol{\mathcal{E}}_0, \textbf{R}, \eta_0)}{\partial \tau_{\kappa, \beta}} \frac{\partial \tau_{\kappa, \beta}}{\partial \mathcal{E}_k}}_{\text{ionic}}.
    \label{equ:partial_derivative_eps}
\end{equation}

Here, $\tau_{\kappa, \beta}$ represents the displacement of atom $\kappa$ along the Cartesian coordinate $\beta$. Evaluating the $\partial \tau_{\kappa, \beta} / \partial \mathcal{E}_k$ expression directly via finite differences is possible, however it goes beyond the scope of this paper. Instead, we expand the ionic contribution in a phonon mode basis and insert Eq.~(\ref{equ:partial_derivative_eps}) into Eq.~(\ref{equ:method:Pockels_E_t_strain}) to obtain

\begin{equation}
    r_{ijk} = \underbrace{-\frac{2}{n_i^2 n_j^2}\chi_{ijk}^{(2)}}_{r_{ijk}^{el}}
    \underbrace{-\frac{1}{n_i^2 n_j^2} \sum_m \frac{\alpha_{ij}^m p_{m,k}}{\omega_m^2}}_{r_{ijk}^{ion}},
    \label{equ:method:r_tot_phonon_basis}
\end{equation}
where $r_{ijk}^{el}$ and $r_{ijk}^{ion}$ are the high-frequency electronic and optical ionic Pockels tensor, respectively. In Eq.~(\ref{equ:method:r_tot_phonon_basis}), $\chi_{ijk}^{(2)}$ refers to the second-order dielectric susceptibility, $n_{i} = \sqrt{\varepsilon_{ii}}$ to the refractive index, $\omega_m$ to the frequency of phonon mode $m$ at the $\Gamma$-point, i.e., in the long wavelength limit, and $\alpha_{ij}^m$ to the Raman susceptibility. The latter can be obtained from 
\begin{equation}
    \alpha_{ij}^m = \sum_{\kappa,\beta} \frac{\partial \chi_{ij}^{(1)}}{\partial \tau_{\kappa, \beta}} u_m(\kappa, \beta),
    \label{equ:raman}
\end{equation}
with $u_m(\kappa, \beta)$ corresponding to the displacement caused by the $m$-th phonon mode and $\chi_{ij}^{(1)}$ to the first-order dielectric susceptibility. Finally, $p_{m,k}$ is the mode polarity
\begin{equation}
    p_{m,k} = \sum_{\kappa, \beta} Z^*_{\kappa, k \beta} u_m(\kappa, \beta)
    \label{equ:mode_polarity}
\end{equation}
where $Z^*_{\kappa,k\beta}$ is the Born effective charge. Throughout this work, the phonon contribution to $r^{ion}$ will further be referred to as $\nu_{ijk} = \sum_m \frac{\alpha_{ij}^m p_{m,k}}{\omega_m^2}$. A detailed derivation leading to Eq.~(\ref{equ:method:r_tot_phonon_basis}) is provided in \cite{veithen2005nonlinear_DFPT}. 

Instead of employing DFPT and the $2n+1$ theorem to compute the $n$, $\chi^{(2)}$, $\alpha$, $p$, $\omega$, and $u$ quantities, we use the \textit{IRamanSpectraWorkChain} framework of Vibroscopy \cite{bastonero2024Vibroscopy}. It relies on DFT and finite differences to calculate the derivatives of the Hellmann-Feynman forces $F$ and polarizations $P$ with respect to atomic displacements and electric fields. The derivatives of $F$ and $P$ provide the force constant matrix $\Tilde{C}$, $Z^*$, $\partial \chi^{(1)}/ \partial \tau$, $\varepsilon$, and $\chi^{(2)}$, from which $\omega$, $u$, $p$, $\alpha$, and $n$ can be extracted, as indicated in Table \ref{tab:Pockels_finite_difference}.

\begin{table}[b]
\begin{ruledtabular}
\begin{tabular}{m{1cm}|ccc}
          & $\partial/\partial \tau$ & $\partial/ \partial \mathcal{E}$ & $\partial^{2}/ \partial^2{\mathcal{E}}$\\
        \hline
        \\[-1em]
       $F$  & $\Tilde{C} \rightarrow \omega, u$ & $Z^{*} \rightarrow p$  & $\partial \chi^{(1)}/ \partial \tau \rightarrow \alpha$ \\
       $P$  & - & $\varepsilon \rightarrow n$ & $\chi^{(2)}$\\
\end{tabular}
\end{ruledtabular}
\caption{\label{tab:Pockels_finite_difference}%
Summary of the Pockels tensor parameters entering Eq.~(\ref{equ:method:r_tot_phonon_basis}). The first row illustrates how the force constant matrix $\Tilde{C}$, the Born effective charge $Z^*$, and the Raman tensor $\partial \chi^{(1)}/ \partial \tau$ can be computed from the derivative of the forces $F$ with respect to atomic displacements $\tau$ and electric fields $\mathcal{E}$. Similarly, the second row illustrates how the high-frequency dielectric tensor $\varepsilon$ and the second-order dielectric susceptibilities $\chi^{(2)}$ are obtained from the derivatives of the polarization $P$. The ``$\rightarrow$'' symbols indicate the related quantities that result from the $F$ and $P$ derivatives and that enter Eq.~(\ref{equ:method:r_tot_phonon_basis}). All derivatives are computed via finite differences with the Vibroscopy software \cite{bastonero2024Vibroscopy}.
}
\end{table}

Forces and polarizations are calculated with the plane-wave DFT code Quantum ESPRES-SO \cite{giannozzi2009quantum_espresso, giannozzi2017quantum_espresso}, overarched by the AiiDA workflow manager \cite{huber2020aiida, uhrin2021aiida_workflows}. Vibrational quantities are obtained via the frozen-phonon approach of phonopy \cite{phonopy-phono3py-JPSJ}, which is embedded in the AiiDA Vibroscopy plugin \cite{bastonero2024Vibroscopy}. Polarizations are also delivered by Vibroscopy, within the framework of the modern theory of polarization \cite{resta1994theoryPolarization} and the electric-enthalpy functional \cite{souza2002finite_electric_field_insulators, umari2002AIMD_finiteDiff_equ_DFPT}. All DFT calculations rely on the Perdew-Burke-Ernzerhof XC-functional for solids (PBEsol) \cite{perdew1996GGA}. It is combined with scalar relativistic, Rappe-Rabe-Kaxiras-Joannopoulos (RRKJ)~\cite{rappe1990rrkjuspp} ultra-soft pseudopotentials with valence configuration $5s^2 6s^2 5p^6 6p^0$ for barium, $3s^2 4s^2 3p^6 3d^0$ for titanium, and $2s^2 2p^4$ for oxygen \cite{dalCorso2014USPP_generation}. 

Our minimal unit cell (``structural prototype'', see Ref. \cite{kotiuga2022structural_prototype}) is made of a \textit{P4bm} supercell (space group 100), constructed from four titanium off-centered primitive cells. The optimized lattice constant is found based on plane-wave and charge density kinetic energy cutoffs of 60 Ry and 600 Ry, respectively, and with a $\Gamma$-centered $4\times 4\times 8$ Monkhorst-Pack \textit{k}-point grid. The same parameters are used for all DFT and DFPT calculations, except when the \textit{k}-point grid density is defined otherwise. The forces and total energy are converged within $\unit{10^{-4}}$ Ry/atom $\approx \unit{1.36 \cdot 10^{-3}}$ eV/atom. For electric field calculations, a finite field step of \unit{5\cdot 10^{-4}} Ry a.u $\approx$ 0.018 \unit{V/\angstrom} is selected and the \textit{k}-point density is doubled along the direction of applied electric field. In frozen-phonon calculations, a $4\times 4\times 4$ supercell, a $2\times 2\times 2$ \textit{k}-point grid, and finite atomic displacements of 0.01 \unit{\angstrom} are employed. Force constants are symmetrized using the translational and the permutation acoustic sum rules. The bandgap and dielectric properties of BTO are corrected with the extended Hubbard model~\cite{campo2010extendedHubbard, van2018extendedHubbard} using the AiiDA Hubbard functionalities~\cite{bastonero2025hubb-aiida}, where onsite \textit{U} and intersite \textit{V} parameters were determined self-consistently via DFPT within linear response, as described in Refs.~\cite{timrov2018hubbard, timrov2021Hubbard_uspp, timrov2022hp}. For this purpose, Löwdin orthogonalized atomic orbitals projectors \cite{lowdin1950hubb_projectors}, the PBEsol XC-functional, and a $2\times 2\times 2$ \textit{q}-point grid were used at fixed cell geometry and fixed atomic positions.

As a final note on nomenclature, the Pockels tensor is contracted based on the Voigt convention \cite{boyd2008nonlinear} when referring to specific coefficients. As a consequence, the three irreducible non-zero entries of the tetragonal Pockels tensor are defined as $r_{13}=r_{113}$, $r_{33}=r_{333}$, and $r_{51}=r_{131} = r_{42}=r_{232}$.

\section{Results and discussion}

Our results are organized in four parts: We first describe how to stabilize the soft phonon modes of tetragonal BTO (Section \ref{chap:resI}). We then use this protocol to compute the Pockels coefficients of this material (Section \ref{chap:resII}). The results, especially the $r_{51}$ coefficient, do not agree well with the experimental data. This issue is resolved in Section \ref{chap:resIV} where the extent of titanium off-centering is modified, resulting in an increasing $r_{51}$ that brings it within the expected experimental range.

\subsection{\label{chap:resI}Dynamic stabilization}
\begin{figure}[h!]
    \centering
    \includegraphics[width=0.5\textwidth]{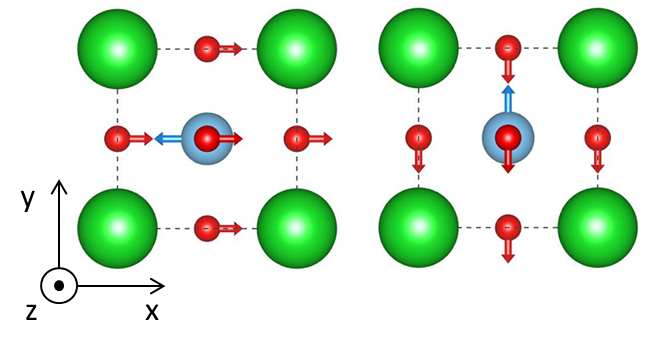}
    \caption{Top view of the $x$- and $y$-polarized soft phonon modes of the BTO \textit{P4mm} unit cell, with barium in green, titanium in blue, and oxygen in red. The eigenvectors of the soft phonon modes, represented by matching color-coded vectors, feature titanium and oxygen out-of-phase oscillations, referred to as Slater-type oscillations.}
    \label{fig:resI:unstable_phonon_modes}
\end{figure}

BTO possesses a tetragonal lattice at room temperature. However, it is rhombohedral at 0 K. A possible static simulation approach involves constraining the cell geometry to a tetragonal structure with \textit{P4mm} symmetry, i.e., a primitive \textit{P} tetragonal cell with a four-fold rotation axis along the $z$-axis and two mirror planes $m$ with crystallographic Miller indices $[200]$ and $[110]$, and reaching static equilibrium. In reality, the \textit{P4mm} high-symmetry phase exhibits two degenerate low-energy soft optical modes, which have been observed experimentally \cite{fontana1994pockels_&_dielectric_relaxation, scalabrin1977phonon_freq_gamma_T, didomenico1968raman_soft_mode} in the lowest frequency range of the vibrational spectrum of BTO, around 1.14 THz. They are characterized by a Slater-type geometry \cite{dwij2020revisiting70y_lattice_dynamics, slater1950lorentz} and their eigenvectors are displayed in Fig.~\ref{fig:resI:unstable_phonon_modes}. However, when computing the phonon bandstructure of the \textit{P4mm} phase of BTO with DFT or DFPT, these modes are found to be degenerate and purely imaginary. This is highly problematic, as both earlier works \cite{veithen2005nonlinear_DFPT, fontana1994pockels_&_dielectric_relaxation, veithen2004Pockels_soft_mode, kim2023Pockels_tetra_BTO} and our results show that they predominantly contribute to the ionic Pockels response of tetragonal BTO. In particular, they drastically affect the largest coefficient $r_{51}$. Thus, without a proper description of the soft phonon modes, the value of $r_{51}$, the coefficient with highest technological relevance, cannot be accurately predicted.

\begin{figure}[h!]
    \centering
    \includegraphics[width=0.4\textwidth]{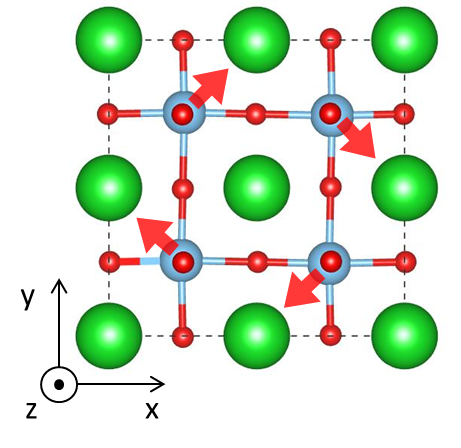}
    \caption{Top view of the BTO \textit{P4bm} $2\times 2\times 1$ structural prototype. Red arrows highlight the in-plane local titanium displacements along the $\langle$110$\rangle$ directions.}
    \label{fig:resI:structural_prototype}
\end{figure}

For a proper description of the structure and dynamics of tetragonal BTO, we use a $2 \times 2 \times 1$ supercell, referred to as structural prototype \cite{kotiuga_2022_privatecomm, kotiuga2022structural_prototype}. Its \textit{P4bm} space group substitutes one mirror plane $m$ for a glide plane $b$, as compared to \textit{P4mm}. More precisely, we now have a primitive tetragonal supercell with a four-fold rotational $z$-axis, a $[220]$ mirror plane, and a $[400]$ glide plane that combines a mirroring action followed by a translation by one half of the supercell lattice parameter. All four titanium atoms are displaced along the $\langle 110 \rangle$ directions, moved along the $x$- and $y$-polarized soft phonon eigenvectors. As a consequence, the local titanium off-centering resembles a vortex-like structure, as shown in Fig.~\ref{fig:resI:structural_prototype}, while preserving the macroscopic spontaneous polarization of tetragonal BTO along $[001]$.

\begin{figure}[h!]
    \centering
    \includegraphics[width=0.7\textwidth]{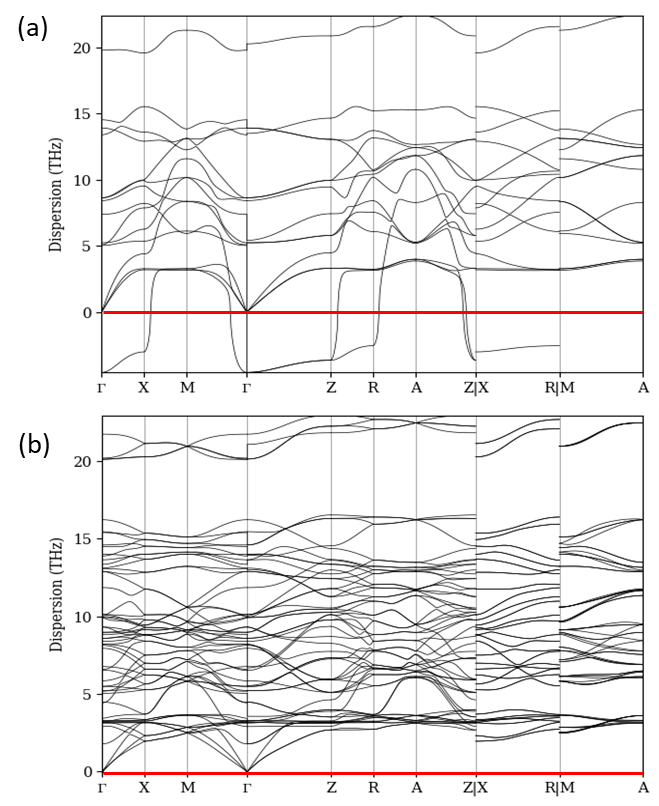}
    \caption{(a) Phonon dispersion of the \textit{P4mm} unit cell exhibiting unstable (negative) branches below the red line. (b) Stable (positive) phonon dispersion of the \textit{P4bm} supercell from Fig.~\ref{fig:resI:structural_prototype}. Both dispersion curves were calculated with the PBEsol XC-functional and RRKJ ultra-soft pseudopotentials. Note that the BZ path is different as cells are different.}
    \label{fig:resI:phonon_dispersion}
\end{figure}

In combination with PBEsol and RRKJ ultra-soft pseudopotentials, the structural prototype allows to stabilize the soft phonon modes. Figures \ref{fig:resI:phonon_dispersion}(a) and (b) confirm this finding, demonstrating that the negative phonon branches pertaining to the \textit{P4mm} unit cell vanish in the \textit{P4bm} structural prototype. Importantly, the latter is not a computational subterfuge to suppress imaginary phonons: It reflects the true structure of tetragonal BTO. X-ray diffraction experiments indeed show the reality of local titanium off-centering. Individual titanium displacements effectively align towards a combination of $\langle 110\rangle$ and $[001]$, the latter referring to the spontaneous polarization displacement along the $c$-axis. More exactly, the titanium displacements were shown to lean away from $[111]$ by 12\unit{\degree} towards the $c$-axis \cite{comes1968BTOtetra_disorder111, stern2004111disorder_12deg}. 

\subsection{\label{chap:resII}Pockels tensor of tetragonal BTO}

\begin{figure}[h!]
    \centering
    \includegraphics[width=0.9\textwidth]{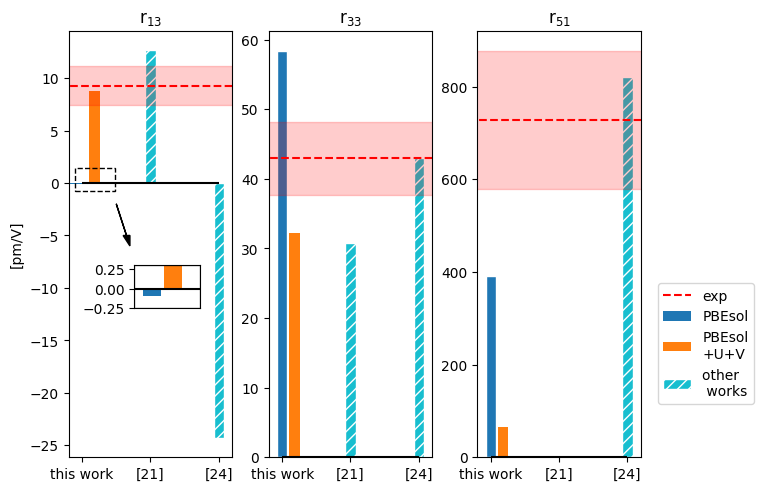}
    \caption{Clamped Pockels coefficients $r_{13}$ (left), $r_{33}$ (center), and $r_{51}$ (right) as calculated with Vibroscopy with PBEsol (dark blue) and PBEsol+\textit{U}+\textit{V} (orange) using the \textit{P4bm} structural prototype of BTO. Other computational works are shown in dashed cyan for comparison. They are based on DFPT, the $2n+1$ theorem, and LDA. Veithen \textit{et al.}~\cite{veithen2005nonlinear_DFPT} used a \textit{P4mm} primitive cell, whereas Kim \textit{et al.}~\cite{kim2023Pockels_tetra_BTO} averaged the responses of four unit cells with different local titanium displacements each. Experimental values and errors are displayed as dashed red lines and shaded areas, respectively \cite{zgonik1994BTOprop}.}
    \label{fig:resII+III:main_results}
\end{figure}

After stabilizing the imaginary phonon modes with the structural prototype of tetragonal BTO, we can calculate its clamped Pockels tensor with the finite-difference DFT approach implemented in Vibroscopy. Against symmetry expectations \cite{gallego2019bilbao}, we obtain a negative and vanishingly small $r_{13}$ coefficient, while a value of 9 \unit{pm/V} is expected from experiments \cite{zgonik1994BTOprop}, as depicted in Fig.~\ref{fig:resII+III:main_results} and summarized in Table \ref{tab:Pockels}. For the second-largest Pockels coefficient $r_{33}$, a value of 58.4 \unit{pm/V} is calculated, which exceeds the experimental data (43 \unit{pm/V}) by 26\%. In contrast, the highest coefficient, $r_{51}$ = 391.2 \unit{pm/V}, severely underestimates the experimental value of 730 \unit{pm/V} by 46\%. 

In an attempt to refine our results, we replace the dielectric tensor with its experimental value (see Table \ref{tab:structural_vibrational_exp_comparison}) in Eq.~(\ref{equ:method:r_tot_phonon_basis}) via the refractive index tensor. This procedure does not significantly change $r_{13}$, which remains small and negative, and leads to an overestimation of $r_{33}$ by 42\%. At the same time, the error on $r_{51}$ decreases from 46\% to 40\%. The values of the experimental dielectric tensor-corrected Pockels coefficients are given in Table \ref{tab:Pockels} under the PBEsol+$\varepsilon_{exp}$ entry. 

Overall, none of our Pockels coefficients computed with PBEsol, with or without dielectric tensor correction, perfectly agrees with other theoretical calculations \cite{veithen2005nonlinear_DFPT, kim2023Pockels_tetra_BTO}, or experimental values \cite{zgonik1994BTOprop}. Kim \textit{et al.} \cite{kim2023Pockels_tetra_BTO} obtained a large negative $r_{13}$ = -24.3 \unit{pm/V}, but $r_{33}$ and $r_{51}$ coefficients matching with experiments. They carried out DFPT calculations of four individual titanium-displaced BTO unit cells with rhombohedral geometry constrained in a tetragonal lattice, using LDA and a corrected dielectric tensor. They averaged the EO response of each of these four unit cells to produce a macroscopic Pockels tensor for tetragonal BTO. On the other hand, the work of Veithen \textit{et al.} \cite{veithen2005nonlinear_DFPT} reported a positive $r_{13} = 12.7$ \unit{pm/V} with accuracy within experimental uncertainty, $r_{33} = 30.8$ \unit{pm/V}, but no $r_{51}$. They relied on DFPT calculations of a high-symmetry \textit{P4mm} unit cell as well as on LDA and a corrected dielectric tensor. Discrepancies between the alternate sign of $r_{13}$ and the (in)ability of calculating $r_{51}$ suggest that both the choice of XC-functional and local titanium off-centering have a significant impact on the Pockels tensor of BTO. The former point is discussed below, whereas the latter will be addressed in Section \ref{chap:resIV}.

The observed discrepancies between our results with PBEsol and those based on LDA in the literature (see Fig.~\ref{fig:resII+III:main_results}) suggest that there might exist another XC-functional that describes the EO properties of tetragonal BTO in a more accurate way than with either LDA or PBEsol. Knowing that correcting the dielectric tensor slightly improves the Pockels coefficient, it appears natural to choose a functional that performs well in reproducing this property. In their study, Kim \textit{et al.} corrected the dielectric tensor by opening the bandgap of BTO with a scissor operator and obtained very good agreement with experimental data for $r_{33}$. Instead of $E_g = 1.88$ \unit{eV} as in our case, they used a corrected value that matched the experimental bandgap $E_g$=3.33 \unit{eV} and corresponds to a 56.5\% increase compared to our calculations with the PBEsol functional. Hence, a functional that corrects both the bandgap and the dielectric tensor might improve the accuracy of Pockels coefficient calculations.

Thanks to the functional-independent capabilities of Vibroscopy, such corrections can be implemented in our workflow in a straightforward manner. Here, we opted for the extended Hubbard model, which removes self-interaction errors by imposing piecewise linearity in the energy functional as a function of occupation~\cite{timrov2018hubbard}. We selected this model because the \textit{U} and \textit{V} parameters can be determined self-consistently with DFPT~\cite{timrov2021Hubbard_uspp} from first-principles, i.e., without prior knowledge of the real bandgap. This is particularly relevant when strain or defects are present in BTO and affect its electronic properties in an unknown manner. A hybrid functional could be used as well to improve the bandgap accuracy, but we discarded this option because of its heavier computational cost. 

\begin{table}[h!]
    \begin{ruledtabular}
    \begin{tabular}{rlccc}
                       & & $r_{13}$ [\unit{pm/V}] & $r_{33}$ [\unit{pm/V}] & $r_{51}$ [\unit{pm/V}] \\
    \hline
    \hline
    Our work             & PBEsol             & -0.1 & 58.4 & 391.2 \\
                         & PBEsol+$\varepsilon_{exp}$  & -0.1 & 61.2 & 436.0\\
                         & PBEsol+\textit{U}+\textit{V}         & 8.8 & 32.2 & 65.5\\
    \hline
    Literature           & \cite{veithen2005nonlinear_DFPT} LDA & 8.9 & 22.3 & - \\
                         & \cite{veithen2005nonlinear_DFPT} LDA + scissor & 12.7 & 30.8 & - \\
                         & \cite{kim2023Pockels_tetra_BTO} LDA + scissor  & -24.3 & 43 & 820.8 \\
                         & \cite{zgonik1994BTOprop} Experiment  & 9 $\pm$ 2 & 43 $\pm$ 5 & 730 $\pm$ 150 \\
    \end{tabular}
    \end{ruledtabular}
    \caption{Clamped Pockels coefficients as computed in this work with PBEsol, PBEsol+$\varepsilon_{exp}$, and PBEsol+\textit{U}+\textit{V} using the \textit{P4bm} structural prototype of tetragonal BTO. Comparison with other works relying on DFPT, the $2n+1$ theorem, with or without a scissor correction of the bandgap, and LDA with a \textit{P4mm} primitive cell \cite{veithen2005nonlinear_DFPT} and an average of four unit cells with local titanium displacements \cite{kim2023Pockels_tetra_BTO}.}
    \label{tab:Pockels}
\end{table}

The extended Hubbard parameters are obtained according to the procedure of \cite{gebreyesus2023HubbardUV_BTO} used for rhombohedral BTO. We apply an onsite \textit{U} correction to the Ti(3d) states and an intersite \textit{V} correction to the Ti(3d)-O(2p) bonds. Through DFPT within linear response theory, we find an optimized values of \textit{U}=5.223 \unit{eV} and a range of \textit{V}=[1.0-1.275] \unit{eV} for each Ti(3d)-O(2p) bonding pair. By applying this correction to the PBEsol structure, the bandgap increases from $E_g = 1.88$ \unit{eV} to $2.92$ \unit{eV}, as compared to standard PBEsol, but it still underestimates the experimental value by 13 \unit{\%} (E$_{g,exp}$ = 3.33 \unit{eV}). However, this correction appears sufficient to improve the error on the dielectric tensor, which falls within 4\% of the experimental data while it is equal to 11\% with standard PBEsol, as reported in Table \ref{tab:structural_vibrational_exp_comparison}. 

\begin{table}[h!]
    \begin{ruledtabular}
    \centering
    \begin{tabular}{rlcccc}
                & & $\omega_{\Gamma}^s$ [THz] & E$_g$ [eV] & $\varepsilon^{\infty}_{11}$ & $\varepsilon^{\infty}_{33}$ \\ \hline \hline
     Our work & PBEsol      & 1.80 & 1.88 & 6.27 & 5.77 \\
     & PBEsol+\textit{U}+\textit{V}  & 3.16 & 2.92 & 5.63 & 5.29 \\
     \hline
     Literature & Experiment & 1.14 & 3.33 & 5.61 & 5.49 \\
    \end{tabular}
    \end{ruledtabular}
    \caption{Soft phonon mode frequency $\omega_{\Gamma}^s$ at the $\Gamma$-point, bandgap energy $E_g$, high-frequency dielectric tensor coefficients $\varepsilon_{11}^{\infty}$ and $\varepsilon_{33}^{\infty}$ using PBEsol and PBEsol+\textit{U}+\textit{V}. The experimental soft phonon mode frequency was obtained from \cite{scalabrin1977phonon_freq_gamma_T}, the bandgap from \cite{wemple1970bandgap}, and the dielectric coefficients from \cite{abel2019largePockels, li1991dielectric}.}
    \label{tab:structural_vibrational_exp_comparison}
\end{table}

When it comes to the Pockels tensor, the inclusion of the \textit{U} and \textit{V} corrections to the PBEsol XC-functional recovers the theoretical tensor symmetry. Furthermore, it seems to improve $r_{13}$ and to a certain extent, $r_{33}$. However, it considerably underestimates $r_{51}$, as shown in Fig.~\ref{fig:resII+III:main_results}. Although $r_{13}$ appears to fall within the experimental range, this result should be considered with precaution. The observed improvement is not caused by the ability of the Hubbard model to describe the electronic properties of BTO more accurately than with the standard PBEsol XC-functional, but by the poor reproduction of the BTO soft phonon modes of the PBEsol+\textit{U}+\textit{V} model. The calculated frequency of these modes, 3.16 \unit{THz}, largely exceeds experimental measurements (1.14 \unit{THz}) and results with PBEsol (1.80 \unit{THz}), as can be seen in Table \ref{tab:structural_vibrational_exp_comparison}). Crucially, the PBEsol+\textit{U}+\textit{V} phonon eigenvectors are fundamentally different from the PBEsol ones, which considerably affects the Raman susceptibilities and mode polarities defined in Eqs.~(\ref{equ:raman}) and (\ref{equ:mode_polarity}), respectively. In particular, the PBEsol+\textit{U}+\textit{V} eigenvectors associated with the soft modes are no longer of the Slater type observed experimentally in \cite{slater1950lorentz}.

\begin{figure}[h!]
    \centering
    \includegraphics[width=0.56\textwidth]{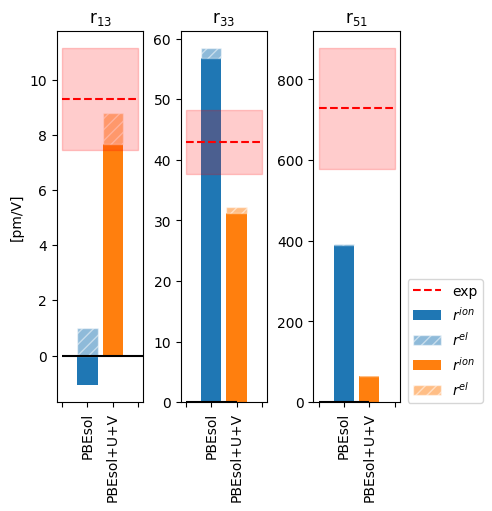}
    \caption{Ionic $r^{ion}$ (solid) and electronic $r^{el}$ (striped) contributions to the clamped Pockels coefficients $r_{13}$, $r_{33}$, and $r_{51}$ calculated with PBEsol (blue) and PBEsol+\textit{U}+\textit{V} (orange). Experimental values and errors are displayed as dashed red lines and shaded areas, respectively \cite{zgonik1994BTOprop}.}
    \label{fig:resII+III:elec_ionic_Pockels}
\end{figure}

The comparison of $r^{el}$ and $r^{ion}$ calculated with PBEsol and PBEsol+\textit{U}+\textit{V} in Fig.~\ref{fig:resII+III:elec_ionic_Pockels} shows that, contrary to expectations, the electronic component ($r^{el}$) of $r_{13}$ is not really affected by the Hubbard correction. The improvement comes from $r^{ion}$, which becomes positive, in the order of 7.8 \unit{pm/V}, while it should not have been impacted by the presence of the \textit{U} and \textit{V} terms. The same issue (variations of $r^{ion}$ instead of $r^{el}$) occurs for $r_{33}$ and $r_{51}$, thus explaining why their value rather deteriorates instead of improving.

\begin{figure}[h!]
    \centering
    \includegraphics[width=1\textwidth]{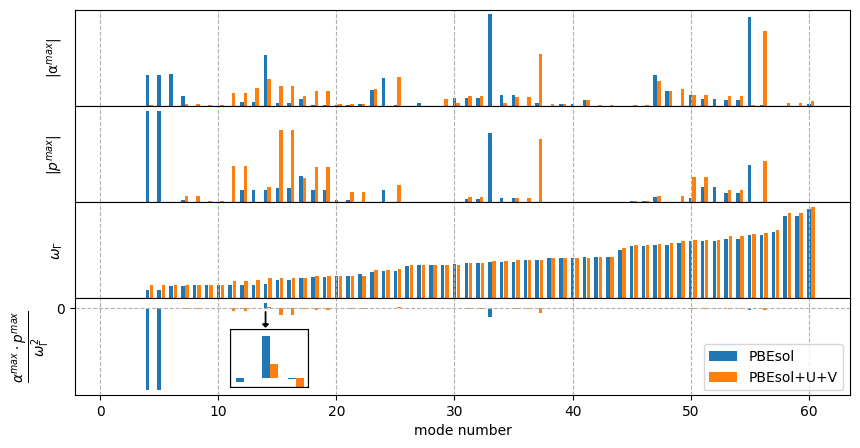}
    \caption{From top to bottom: maximum Raman susceptibility tensor coefficient $max|\alpha_{ij}|$, maximum mode polarity coefficient $max|p_k|$, phonon frequency $\omega_{\Gamma}$, and maximum phonon contribution $\nu^{max} = \alpha^{max} p^{max}{\omega^{-2}_{\Gamma}}$ as a function of the phonon mode index $m$ using PBEsol (blue) and PBEsol+\textit{U}+\textit{V} (orange).}
    \label{fig:resII+III:mode_contribution}
\end{figure}

To understand the origin of the $r^{ion}$ variability with respect to the XC-functional, we further examine the phonon contribution to $r^{ion}$, i.e., $\nu_{ijk}$, by excluding the dielectric tensor renormalization. We decompose $\nu_{ijk}$ into a product of Raman susceptibilities, mode polarities, and inverse soft phonon frequencies and report these quantities in Fig.~\ref{fig:resII+III:mode_contribution} as a function of the phonon mode index. Since the Raman susceptibility tensors $\alpha_{ij}^m$ are $3\times 3\times m$-dimensional objects, we define $|\alpha^{max}|$ as the largest $ij$-th tensor coefficient for each mode index $m$. The same applies to the mode polarities $p_{m,k}$ with $m \times 3$ dimensionality, where we set $|p^{max}|$ as the maximum $k$-th coefficient for each $m$. By doing so, we can highlight the upper bound discrepancies between the PBEsol and PBEsol+\textit{U}+\textit{V} calculations. 

Figure~\ref{fig:resII+III:mode_contribution} clearly indicates that $|\alpha^{max}|$ and $|p^{max}|$ are fundamentally different with and without the Hubbard correction, especially for the soft modes with indices 4 and 5. PBEsol is able to capture their large Raman susceptibilities and large mode polarities \cite{hamze2020design_rules_Pockels}, while PBEsol+\textit{U}+\textit{V} is not. This capability is particularly critical since the largest Pockels coefficient of tetragonal BTO is mainly related to the activity of the soft phonon modes, as demonstrated by calculating the maximal phonon contribution to $r^{ion}$, $\nu^{max} = \alpha^{max} p^{max}{\omega^{-2}_{\Gamma}}$, where $\nu^{max}$ is dominated by modes 4 and 5. Conversely, $\nu^{max}$ calculated with PBEsol+\textit{U}+\textit{V} does not feature these essential mode contributions. It follows that the value for $\nu_{113}$, and, in turn, $r_{13}$ with PBEsol+\textit{U}+\textit{V} results from a number of small, yet inaccurate phonon contributions that sum up to an ostensibly better result without any physical justification. 

Based on this finding, we conclude that the Hubbard model is unsuitable to improve the physics of the Pockels tensor for BTO when superseding the PBEsol XC-functional. Reference~\cite{gebreyesus2023HubbardUV_BTO} supports our conclusion through a comprehensive analysis of the role of the extended Hubbard correction for rhombohedral BTO. There, it is shown that phonon and Raman properties critically depend on the crystal geometry, whose description is more accurate with PBEsol than with the Hubbard model. The strength of PBEsol+\textit{U}+\textit{V} is that it returns a more accurate bandgap and dielectric tensor than PBEsol. In other words, we only expect the electronic contributions to $r^{el}$ and $r^{ion}$ to improve upon Hubbard corrections, albeit to a limited extent as argued earlier when replacing the dielectric tensor with its experimental value (see Table \ref{tab:Pockels}).

In reality, BTO thin-films are affected by point and extended defects, as well as by the presence of ferroelectric domains \cite{abel2013strongPockels, winiger2024exp_pockels}. Hence, the experimental Pockels coefficients of real BTO monocrystals are expected to be smaller than those calculated from DFT on a pristine cell. However, quantification of these non-ideal factors goes beyond the scope of this paper. Similarly, it should be mentioned that the influence of other factors such as the vibrational activity or finite temperature effects, should not be ruled out \textit{a priori}. As discussed in Section \ref{chap:resI}, $r_{51}$ has a predominant vibrational character, driven by soft phonon modes. In addition, the standard DFT is a 0 K theory, neglecting the thermal excitations of the nuclei \cite{monacelli2021SSCHA}. Thus, both effects could affect the value of the Pockels coefficients, as demonstrated by Veithen \textit{et al.}~\cite{veithen2005nonlinear_DFPT}. In a subsequent publication \cite{veithen2005temperature}, the same authors calculated $r_{51}$ as a function of temperature. They used an effective Hamiltonian and a low-order Taylor expansion of the energy around the paraelectric phase, with ionic degrees of freedom aligned with the unstable phonon eigenvectors. In this configuration, they obtained an unclamped $r_{51}$ of 622 \unit{pm/V} at room temperature, including piezoelectric contributions. While encouraging, these results severely underestimate the experimental unclamped value of 1500 \unit{pm/V} when piezoelectric contributions are taken into account. Finally, the Pockels response is a frequency-dependent quantity~\cite{winiger2024exp_pockels}, as the dielectric response. However, in this work, we do not consider these frequency dependencies because they are computationally too expensive to account for at the \textit{ab initio} level. We sample Pockels responses at lattice (phonons) and electronic resonances, and sum up individual contributions. This approach has the advantage of being less intensive, and is valid as long as the electric field modulation frequency is smaller than the lowest optical phonon mode frequency (i.e., 1.14 THz, see Table \ref{tab:structural_vibrational_exp_comparison}). This is the case in high-speed, GHz electro-optical modulators~\cite{abel2019largePockels}.

\subsection{\label{chap:resIV}Pockels coefficients and local titanium off-centering}

By replacing the dielectric tensor with its experimental value, we still underestimate the experimental $r_{51}$ by a factor of 1.6 (436 vs. 700 \unit{pm/V}). Correcting first-order dielectric susceptibilities is thus insufficient. As discussed in Section \ref{chap:resI}, the vibrational properties, and in particular the soft phonon modes, are very sensitive to local titanium off-centering. As a consequence, Veithen \textit{et al.}~\cite{veithen2005nonlinear_DFPT} could not report any $r_{51}$ using the high-symmetry \textit{P4mm} structure and the LDA XC-functional because of the presence of imaginary phonon modes. The inadequacy of this structure was further confirmed by Kim \textit{et al.}~\cite{kim2023Pockels_tetra_BTO}, who stressed that a cell with local titanium disorder mediated by unstable phonon modes is required to calculate $r_{51}$. Although we use such a structure, our results do not agree with experiments nor with Kim \textit{et al.}.

\begin{figure}[h!]
    \centering
    \includegraphics[width=0.4\textwidth]{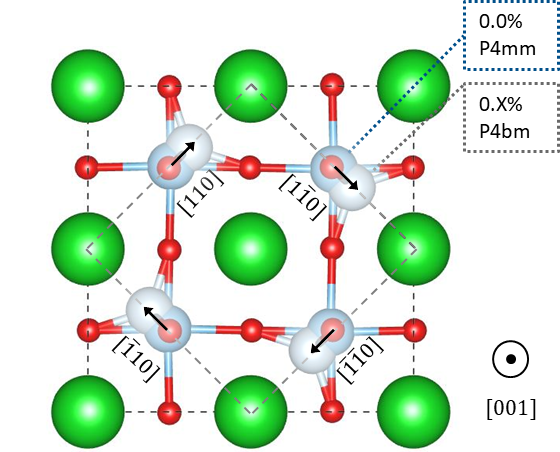}
    \caption{Top view of the tetragonal BTO structural prototype with barium in green, titanium in light blue, and oxygen in red. The high-symmetry titanium positions corresponding to the \textit{P4mm} geometry are labeled with 0.0\%. 0.X\% indicates the magnitude of the displacements of the titanium atoms along the $\langle 110 \rangle$ directions, marked by black arrows. As a reference, a 100\% displacement means that the titanium atoms occupy the same positions as the barium ones and were moved by a distance corresponding to $a\frac{\sqrt{2}}{4}$, $a$ being the $2\times 2\times 1$ supercell lattice constant. The barium and oxygen atoms are fixed to the \textit{P4bm} optimized geometry positions. }
    \label{fig:resIV:Ti_displacement}
\end{figure}

One parameter that has not yet been explored in our study is the degree of titanium off-centering applied to the \textit{P4bm} supercell and how it affects the Pockels tensor. To shed light on this phenomenon, we created a series of \textit{P4bm} structures, where titanium was incrementally displaced along the $\langle 110 \rangle$ directions, as illustrated in Fig.~\ref{fig:resIV:Ti_displacement}. A 0.0\% (0.466\%) displacement of titanium atoms along the $\langle 110 \rangle$ diagonals of the $2\times 2\times 1$ supercell corresponds to the \textit{P4mm}, high-symmetry structure (structural prototype shown in Fig.~\ref{fig:resI:structural_prototype}), while 100\% means that the positions of the titanium atoms coincide with those of the barium ones.

\begin{figure}[h!]
    \centering
    \includegraphics[width=0.8\textwidth]{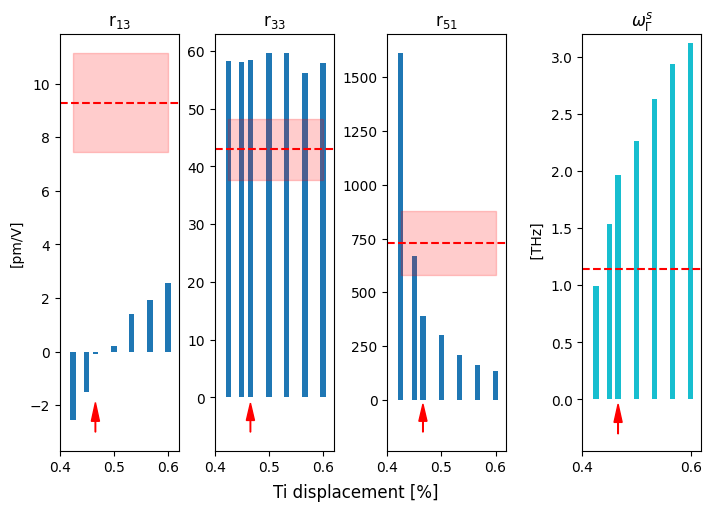}
    \caption{Clamped Pockels coefficients $r_{31}$ (left), $r_{33}$ (center left), $r_{51}$ (center right), and soft phonon frequencies $\omega_{\Gamma}^s$ (right) as a function of titanium displacements along the $\langle 110 \rangle$ directions (see Fig.~\ref{fig:resIV:Ti_displacement}) computed with PBEsol. The titanium displacements corresponding to the \textit{P4bm} ground-state geometry are indicated by the red arrows at 0.466\%. Experimental values are displayed as dashed red lines encompassed by their uncertainty ranges in shaded red \cite{zgonik1994BTOprop}.}
    \label{fig:resIV:main_results}
\end{figure}

Following the methodology of Section \ref{chap:resII}, we recalculated $r_{13}$, $r_{33}$, and $r_{51}$ with PBEsol for a series of \textit{P4bm} titanium-displaced structures around the ground-state (0.466\% displacement). Results are shown in Fig.~\ref{fig:resIV:main_results}, where the titanium displacements associated with the \textit{P4bm} supercell ground-state are indicated by the red arrows. We observe that the smallest coefficient, $r_{13}$, becomes positive with increasing titanium off-centering. In contrast, it remains negative for geometries between the \textit{P4mm} (0\% titanium displacement) and the \textit{P4bm} ground-state. Yet, all three coefficients, $r_{13}$, $r_{33}$, and $r_{51}$, should have the same sign according to symmetry rules \cite{gallego2019bilbao}. It turns out that phonon mode number 14 contributes positively to $\nu^{max}$, as magnified in Fig.~\ref{fig:resII+III:mode_contribution}, leading to an overall negative $r^{ion}$ at 0.466\% displacement. As titanium off-centering increases, the magnitude of $\nu^{max}$ related to mode 14 decreases. As a result, $r_{13}$ becomes increasingly positive.

The titanium-displaced calculations provide us with insightful information about $r_{33}$. It appears to be the least affected coefficient by the titanium off-centering. In fact, the distribution of $r_{33}$ does not change by more than 1.3 \unit{pm/V} around its mean when the titanium  atoms are displaced. This indicates that neither the soft phonon modes nor mode 14 affect the value of $r_{33}$. Since in the $r_{33}$ configuration of BTO-based MZI, the incoming light, the outgoing light, and the electric field are all polarized along the $c$-axis (indexed by $i=j=k=3$), none of them interacts with in-plane $x$- or $y$-polarized modes.

Finally, we thoroughly examine the behavior of $r_{51}$, since it is the most relevant coefficient for practical applications, with an experimental clamped Pockels response of 730 $\pm$ 150 \unit{pm/V}. We demonstrate that, starting from the \textit{P4bm} ground-state at 0.466\% displacement, moving the titanium positions towards the \textit{P4mm} geometry along the $\langle 110 \rangle$ directions considerably increases $r_{51}$. It can be seen in Fig.~\ref{fig:resIV:main_results} that a 0.45\% titanium displacement brings $r_{51}$ within the experimental range (see Table \ref{tab:Pockels_Tidisp}), while 0.425\% titanium displacements boost it to 1614.6 \unit{pm/V}. At the same time, the frequency of the soft phonon modes decreases from 2 \unit{THz} at 0.466\% down to 1.5 (1.0) \unit{THz} at 0.45\% (0.425\%). The experimental value at room temperature is 1.13 \unit{THz} \cite{scalabrin1977phonon_freq_gamma_T}. Overall, the displacement-dependent results indicate that a structure slightly different from the \textit{P4bm} ground-state, with a shift of the titanium atoms along the $\langle$110$\rangle$ direction by 0.425 to 0.45\%, provides accurate $r_{33}$ and $r_{51}$ results as well as a close-to-experiment soft phonon frequency. The fact that $r_{13}$ is negative in such a configuration suggests that we might be missing some physical contribution to the calculation of this coefficient. Nonetheless, the magnitude of $r_{13}$ is much smaller  than that of $r_{33}$ or $r_{51}$, and thus less significant for technological applications.

\begin{table}[h!]
    \begin{ruledtabular}
    \begin{tabular}{lcccccc}
                       & $r_{13}$ [\unit{pm/V}] & $r_{33}$ [\unit{pm/V}] & $r_{51}$ [\unit{pm/V}] & $\omega_{\Gamma}^s$ [\unit{THz}] & $\varepsilon_{11}$ & $\varepsilon_{33}$\\
    \hline
    \hline
    PBEsol @ 0.425\% & -2.5 & 58.3 & 1614.7 & 1.0 & 6.12 & 5.62\\
    PBEsol @ 0.45\%  & -1.5 & 58.0 & 667.0 & 1.5 & 6.11 & 5.62\\
    PBEsol @ 0.466\% & -0.1 & 58.4 & 391.2 & 2.0 & 6.10 & 5.62\\
    \end{tabular}
    \end{ruledtabular}
    \caption{Clamped Pockels coefficients of tetragonal BTO, soft phonon frequency and dielectric coefficients as computed with PBEsol, using the $2\times 2\times 1$ structural prototype with 0.425\%, 0.45\%, and 0.466\% titanium displacements along the $\langle 110 \rangle$ directions. The 0.466\%-labeled structure corresponds to the ground-state of \textit{P4bm} BTO.}
    \label{tab:Pockels_Tidisp}
\end{table}

The \textit{P4bm} structure can also be obtained by relaxing the atomic positions from the high-symmetry \textit{P4mm} unit cell along the four possible directions induced by the soft phonon eigenvectors and then building a supercell combination thereof. This observation potentially explains the findings of Kim \textit{et al.}~\cite{kim2023Pockels_tetra_BTO} in Fig.~\ref{fig:resII+III:main_results}. They obtained a larger $r_{51}$ than ours with 824 \unit{pm/V} by averaging the response of four different unit cells, which they constructed by relaxing the atomic positions along the unstable phonon eigenvector directions. Each of their unit cells is characterized by a titanium displacement of 0.7\% with respect to the unit cell $\langle 110 \rangle$ directions. This percentage must be divided by two to be compared with ours since our reference structure is twice as wide. This translates into a 0.35\% titanium displacement, which is (much) smaller than the 0.466\% titanium displacement of the \textit{P4bm} ground-state and whose $r_{51}$ = 1844 \unit{pm/V}~\cite{kim2023Pockels_tetra_BTO} is even smaller than our result with 0.425\% ($r_{51}$ = 1614.7 \unit{pm/V}). Comparing our results with the values of Kim \textit{et al.} reveals the critical sensitivity of the largest Pockels coefficient with respect to the crystal structure.

Conceptually, decreasing titanium off-centering corresponds to a rotation of the spontaneous polarization away from $[111]$ and towards $[001]$. The authors of \cite{ong2012strain} used first-principles calculations to show that decreasing titanium off-centering is associated with a change of crystalline phase, i.e., from paraelectric cubic to tetragonal, and in consequence with a divergence of the in-plane dielectric constant. In addition, the work of \cite{fredrickson2018BTO_strain} concluded that a diverging dielectric tensor causes the Pockels response to increase, together with a softening of the low-energy optical phonon modes. These results are in agreement with our findings: We observe an increase in the in-plane dielectric constant $\varepsilon_{11}$, a decrease in the soft phonon mode frequency $\omega_{\Gamma}^s$, and an increase in the $r_{51}$ Pockels coefficient with decreasing titanium off-centering (see Table \ref{tab:Pockels_Tidisp}).

\section{Data availability statement}
The data that support the findings of this article are openly available on the Materials Cloud Archive~\cite{data_archive}.

\section{\label{chap:conclusion}Conclusion}

In this study, we present an automated \textit{ab initio} computational framework to calculate the clamped Pockels tensor of tetragonal BTO for any XC-functional. Our method relies on DFT, Hellmann-Feynman forces, the modern theory of polarization, the electric-enthalpy functional, and finite-difference derivatives to compute up to second-order dielectric and vibrational contributions to the Pockels tensor.

We obtained a stable tetragonal phase for BTO by combining a \textit{P4bm} supercell prototype with local titanium off-centering, the PBEsol XC-functional, and ultra-soft pseudopotentials. As a result, we recovered the soft phonon modes that are responsible for the large $r_{51}$ Pockels coefficient of BTO. We calculated the non-zero tensor coefficients, $r_{13}$, $r_{33}$, and $r_{51}$ and determined that $r_{13}$ is vanishing small, while overestimating $r_{33}$ by 35\% and underestimating $r_{51}$ by 46\%. Correcting the dielectric tensor with its experimental value did not improve $r_{51}$, the coefficient of highest technological relevance. Ultimately, we found out that PBEsol is the most suitable XC-functional to calculate the Pockels tensor of tetragonal BTO.

One of the key findings of this paper is that local titanium off-centering strongly impacts the $r_{51}$ response. We showed that reducing the titanium off-centering considerably increases $r_{51}$, provided that the soft phonon modes frequencies at the $\Gamma$-point remain positive. Following this procedure, we successfully obtained $r_{51}$ within 4\% of the experimental value at 0.45\% displacement of the titanium ions along $\langle 110 \rangle$.

Our calculation workflow, which is implemented in the open-source Vibroscopy package and AiiDA-plugin, is freely available \cite{bastonero2024Vibroscopy}. It is capable of handling any XC-functional implemented in the Quantum ESPRESSO code to accurately compute the Pockels tensor of materials. For example, it will be possible to investigate the influence of defects on the EO properties of BTO, which is of practical relevance to design and optimize Pockels-based devices.

\begin{acknowledgments}
We would like to thank (in alphabetical order) Luiz Felipe Aguinsky, Marnik Bercx, Nicola Spaldin, Iurii Timrov, and Andrea Urru for their valuable insights and technical support. This project was funded by Innosuisse, the Swiss Innovation Agency, under project number 102.262 IP-ICT. Computer time was provided by the Swiss National Supercomputing Centre (CSCS) under projects s1119 and lp16.
L.~B. and N.~M. gratefully acknowledge support from the Deutsche Forschungsgemeinschaft (DFG) under Germany's Excellence Strategy (EXC 2077, No. 390741603, University Allowance, University of Bremen) and Lucio Colombi Ciacchi, the host of the ``U Bremen Excellence Chair Program''.
N.~M acknowledges support from the NCCR MARVEL, a National Centre of Competence in Research, funded by the Swiss National Science Foundation (grant number 205602).
\end{acknowledgments}


\bibliography{article}

\end{document}